\documentclass{article}

\usepackage{arxiv}
\usepackage{amsmath, amsfonts, amssymb}
\usepackage{times}
\usepackage{slashed}
\usepackage{tabularx}
\usepackage{multirow}
\usepackage[utf8]{inputenc} % allow utf-8 input
\usepackage[T1]{fontenc}    % use 8-bit T1 fonts
\usepackage{hyperref}       % hyperlinks
\usepackage{url}            % simple URL typesetting
\usepackage{booktabs}       % professional-quality tables
\usepackage{amsfonts}       % blackboard math symbols
\usepackage{nicefrac}       % compact symbols for 1/2, etc.
\usepackage{microtype}      % microtypography
\usepackage{lipsum}		% Can be removed after putting your text content
\usepackage{graphicx}
\usepackage{siunitx}

\usepackage{braket}
\usepackage{xcolor}
\definecolor{frenchblue}{rgb}{0.0, 0.45, 0.73}
\definecolor{RoyalRed}{RGB}{157, 16, 45}

\hypersetup{
  linkcolor  = frenchblue,
  urlcolor    = frenchblue,
  citecolor  = frenchblue,
  colorlinks = true,
  pdftitle     = {CP-violation in the spin-1/2 singly charmed baryons},
  pdfauthor = {Y.~ \"Unal, Ulf-G. ~Mei{\ss}ner },
  pdfkeywords = {},
}

\makeatletter
\renewcommand{\maketitle}{\bgroup\setlength{\parindent}{0pt}
\begin{flushleft}
  \textbf{\@title}
  \@author
\end{flushleft}\egroup
}
\makeatother

\title{\Large{Strong CP violation in spin-1/2 singly charmed baryons} \\ }

\author{ \vspace*{10pt}
    \textbf{Y.~\"Unal}$^{1,2,a}$, 	\textbf{Ulf-G.~Mei{\ss}ner}$^{1, 3, 4,b}$\ \\ 
	$^{1}$ \textit{Helmholtz-Institut f\"ur Strahlen- und Kernphysik and Bethe Center for Theoretical Physics 
	Universit\"at Bonn, D-53115 Bonn, Germany} \\
	$^{2}$\textit{Physics Department, \c{C}anakkale  Onsekiz Mart University 
	17100 \c{C}anakkale, Turkey} \\
	$^{3}$\textit{Institute for Advanced Simulation, Institut f\"ur Kernphysik and 
    J\"ulich Center for Hadron Physics, Forschungszentrum J\"ulich, D-52425 J\"ulich, Germany} \\
    $^{4}$\textit{Tbilisi State University, 0186 Tbilisi, Georgia} \\
	$^{a}$\href{mailto:unal@hiskp.uni-bonn.de} {unal@hiskp.uni-bonn.de},
	$^{b}$\href{mailto:meissner@hiskp.uni-bonn.de}{meissner@hiskp.uni-bonn.de	}	
}
\begin{document}

\clearpage
\maketitle
\thispagestyle{empty}

\textbf{\textit{Abstract}}~  We report on the calculation of the  CP-violating form factor $F_3$ and the
  corresponding electric dipole moment for charmed
  baryons in the spin-1/2 sector generated by the QCD $\theta$-term.
  We work  in the framework of  covariant baryon chiral perturbation theory within the
  extended-on-mass-shell renormalization scheme up to next-to-leading order in the chiral
  expansion.


\section{Introduction}

Charge conjugation and parity symmetry (CP) violation is an essential condition for an
asymmetry between matter and anti-matter in the present universe. On the other hand, CP
violation by the complex phase of the Cabibbo-Kobayashi-Maskawa (CKM) quark-mixing matrix is
insufficient to explain the dominance of matter over antimatter \cite{Gavela:1993ts, Huet:1994jb},
meaning that the presence of other CP violating mechanisms within or beyond the Standard Model (SM)
is required.
The QCD $\theta$-term is the only source of P and T violation within the SM beyond the complex phase
of the CKM quark-mixing matrix. However, because of the significant suppression of electric dipole
moments (EDMs) induced by the complex phase of the CKM matrix and unobservably small CKM backgrounds,
any measurement of EDM of any quantum system would indicate of presence CP violation beyond the
CKM mechanism in the SM. EDMs are important observables generated by the CP-violating effects.
Thus, measurements of hadron EDMs lead to severe restrictions in the mechanim generating CP violation,
as detailed e.g. in Ref.~\cite{Dekens:2014jka}.

CP violation has recently been established in the charm sector, more precisely in the 
meson decays $D^0 \to K^-K^+$ and $D^0 \to \pi^-\pi^+$~\cite{Aaij:2019kcg}, and
LHCb has also measured the difference of  CP-asymmetry of the three-body singly
Cabibbo-suppressed $\Lambda_c^+$ decays~\cite{Aaij:2017xva}. There have also
been quite a number of studies predicting CP asymmetries in charmed baryon decays,
see e.g.~\cite{Grossman:2018ptn} and references therein. It is therefore of interest to
investigate other possible effects of CP violation in singly-charmed baryons.
Indeed, a first measurement of CP violation in $\Xi_c^+\to pK^-\pi^+$ decays
has been performed by LHCb~\cite{Aaij:2020wil}. However, these data are consistent
with the hypothesis of no CP violation. On the other hand, another experimental study
promises to search for direct CP violation by measuring the asymmetries of three different
decay channels of the $\Lambda_c^+$ baryon ~\cite{Shi:2019vus}. 

Here, we concentrate on the effects generated
by the strong CP-violating $\theta$-term of QCD, that also induces electric dipole
moments in light baryons, as pioneered in Refs.~\cite{Baluni:1978rf,Crewther:1979pi}.
The proper framework to address such questions is baryon chiral perturbation theory,
see~ \cite{Bernard:2007zu} for a review.
In fact, the masses, axial charges, and electromagnetic decays of the
charmed and bottomed baryons have already been calculated in the framework
of the  heavy baryon approach~\cite{Jiang:2014ena,Jiang:2015xqa}. More recently,
the magnetic moments of the spin-1/2 singly charmed baryons were analyzed in covariant
baryon chiral perturbation theory~\cite{Xiabng:2018qsd}. In this paper, we extend these
studies and work out the CP-violating effects induced by the QCD $\theta$-term.
While there is experimental activity to assess the effects of
the $\theta$-term in neutron and proton EDMs, singly-charmed baryons offer
a completely new venue towards these elusive effects, with very different
systematic uncertainties that hamper such measurements. How competitive
these measurements will be would require a much more refined analysis as
presented here.
We note that recent progress towards the first measurement of the charm baryon
dipole moments has been reported in Ref.~\cite{Aiola:2020yam}, thus our investigation
is very timely. In the near future, further measurements with charmed hadrons, along with
different theoretical improvements, would help to further elucidate the CP violation
in the charm quark sector.

The manuscript is organized as follows. In Section~\ref{sec:Lagr}, we briefly
discuss the underlying chiral Lagrangian. The CP-violating electromagnetic form factor
of the singly-charmed baryons is worked out in Section~\ref{sec:ff} followed by the
display of our numerical results in Section~\ref{sec:res}. Section~\ref{sec:summ}
contains the summary and outlook. The appendices contain some technicalities as
well as more detailed tables of results.

\section{Chiral Lagrangian including CP-violating terms}
\label{sec:Lagr}
The QCD Lagrangian of the strong interactions including the $\theta$ term reads
\begin{equation}
  \mathcal{L}_{\text{QCD}}=\bar{q}(i\slashed{D}-\mathcal{M})q-\frac{1}{4}\mathcal{G}_{a\mu \nu}
  \mathcal{G}_a^{\mu\nu}+\frac{g^2 \theta}{64 {\pi}^2}\varepsilon_{\mu \nu \rho \sigma}
  \mathcal{G}_{a}^{\mu \nu} \mathcal{G}_a^{\rho\sigma}, \quad \quad a = 1, ..., 8~, 
\end{equation}
where $ \mathcal{G}_{a}^{\mu \nu}$ is the gluon field-strength tensor, $g$ is the strong
coupling constant and $\mathcal{M}$ is
the quark mass matrix. Strong CP violation arising from the U(1) anomaly in QCD is specified
via the vacuum angle $\theta$. The measurable quantity is not $\theta$ but
the combination
\begin{equation}
\theta_0= \theta + \text{arg det} \mathcal{M},
\end{equation}
because of the anomaly. Here, to describe the phenomena related to the $\theta$-term,
we seek a description in a properly tailored effective field theory, see e.g.
Refs.~\cite{deVries:2012ab,Bsaisou:2014oka}
for the detailed construction of the corresponding effective Lagrangian to one loop accuracy.

The Goldstone bosons together with the flavor singlet $\eta_0$,
resulting from the spontaneous symmetry breaking of $U(3)_R \times U(3)_L$ into $U(1)_V$,
are represented by the matrix-valued field $\tilde{U}$. Treating the vacuum angle
$\theta(x)$ as an external field, it transforms as $\theta (x) \rightarrow \theta'(x)=
\theta(x)-2 N_f \alpha$ under axial U(1) rotations, with $N_f$ the number of flavors,
and $\alpha$ is the rotation angle. Following the spontaneous chiral symmetry
breaking, under the axial U(1) transformation, $\tilde{U}$ changes but the combination
of $\bar{\theta}_0(x)=\theta(x)-i ~ \text{ln} ~ \text{det} ~ \tilde{U}(x)$ stays invariant.
Using this invariant combination of $\bar{\theta}_0(x)$, one can construct the most general
mesonic chiral effective Lagrangian up-to-and-including second chiral order
\begin{equation}
\begin{aligned}
  \mathcal{L} =& -V_0+V_1\langle\nabla_\mu \tilde{U}^\dagger \nabla^\mu \tilde{U} \rangle+V_2
  \langle \tilde{\chi} \tilde{U}+\tilde{\chi} \tilde{U}^\dagger \rangle+i V_3 \langle \tilde{\chi}
  \tilde{U} - \tilde{\chi} \tilde{U}^\dagger \rangle\\		
  &+V_4 \langle \tilde{U} \nabla_\mu \tilde{U}^\dagger\rangle
  \langle \tilde{U}^\dagger \nabla^\mu \tilde{U} \rangle.
\end{aligned}
\label{Meslagr}
\end{equation}
 %+V_5 \langle \nabla_\mu \theta \nabla^\mu \theta \rangle%
Note that $\langle...\rangle$ denotes the trace in flavor space and $\tilde{\chi} =
2B_0\mathcal{M}$, with the light quark mass matrix $\mathcal{M} =\text{diag} (m_u, m_d, m_s)$.
The covariant derivative of $\tilde{U}$ is given by
\begin{equation}
\nabla_\mu \tilde{U} = \partial_\mu \tilde{U}-i (v_{\mu}+a_{\mu}) \tilde{U}+i\tilde{U}(v_{\mu}-a_{\mu}),
\end{equation}
where $v_{\mu}$ and $a_{\mu}$ are the conventional vector and axial-vector external sources.
The $V_i$ coefficients in the Lagrangian~\eqref{Meslagr} are functions of $\bar{\theta}_{0}$. 
One needs to determine the vacuum expectation value of $\tilde{U}$ in order to include non-trivial
vacuum effects based on the angle $\theta_0$. Parameterizing the vacuum as
\begin{equation}
U_0=\text{diag}(e^{-i\varphi_u}, e^{-i\varphi_d}, e^{-i\varphi_s}),
\end{equation}
the minimized potential energy $V(U_0)$ can be determined using the notation $\bar{\theta}_0
=\theta_0-\sum_{q} \varphi_{q}$. In this way, the Taylor expansions of the $V_i$ functions in terms of
$\bar{\theta}_0$ yield

\begin{equation}
\begin{aligned}
V_i( \bar{\theta}_0)=& \sum_{n=0}^{\infty}V_i^{(2n)} \bar{\theta}_0^{2n} \  \quad\quad \text{for} \quad i=0, 1, 2, 4 \\
V_3( \bar{\theta}_0)=& \sum_{n=0}^{\infty}V_i^{(2n+1)} \bar{\theta}_0^{2n+1}. \
\end{aligned}
\end{equation}

Note that while all other $V_i$ are even function of $\bar{\theta}_0$, $V_3$ is odd.  To express the Lagrangian
in terms of the angles $\varphi_q$ one then writes the $\tilde{U}$ with the vacuum expectation
value $U_0$ as $\tilde{U}=\sqrt{U_0}U\sqrt{U_0}$ by choosing
\begin{equation}
U=\text{exp}\left(i \sqrt{\frac{2}{3}} \frac{\eta_0}{F_0}+i \frac{\sqrt{2}}{F_{\pi}} \phi\right),
\end{equation} 
where $\phi$ represents the Goldstone boson octet
\[
  \phi=
  \begin{pmatrix}
               \frac{1}{\sqrt{2}}\pi^0+ \frac{1}{\sqrt{6}}\eta_8                      & \pi^+        & K^+ \\
               \pi^-      &        -\frac{1}{\sqrt{2}}\pi^0+ \frac{1}{\sqrt{6}}\eta_8                        & K^0 \\
    K^-                & \bar{K}^0                &     - \frac{2}{\sqrt{6}}\eta_8         
  \end{pmatrix}.
\]         
Thus, the chiral effective Lagrangian in terms of the Goldstone boson fields composed in $\tilde{U}$
reads \cite{Borasoy:2000pq}
\begin{equation}
\begin{aligned}
  \mathcal{L}_{\phi}=  ~&-V_0+V_1\langle\nabla_\mu U^\dagger \nabla^\mu U \rangle
  +(V_2+\mathcal{B}V_3)\langle	\chi(U+U^\dagger)\rangle-i \mathcal{A}V_2 \langle U-U^\dagger
  \rangle \\		
  &+\mathcal{A}V_3 \langle U+U^\dagger \rangle+V_4 \langle U \nabla_\mu U^\dagger\rangle
  \langle U^\dagger \nabla^\mu U \rangle~.
\end{aligned}
\end{equation}
Here $\chi=~2B_0 \text{diag}(m_u \text{cos}
\varphi_u, m_d \text{cos} \varphi_d, m_s \text{cos} \varphi_s)$. To leading order, $\mathcal{A}$ and $\mathcal{B}$ are given as
\begin{eqnarray}
\mathcal{A}=\frac{V_0^{(2)}}{V_2^{(0)}} \bar{\theta}_0 + \mathcal{O}({\delta}^4), \quad \quad \mathcal{B}= \frac{V_3^{(1)}}{V_2^{(0)}} \bar{\theta}_0 + \mathcal{O}({\delta}^6).
\end{eqnarray}
After vacuum alignment,  the $V_i$ coefficients are now functions of $\bar{\theta}_0+ \sqrt{6}
\eta_0/F_0$. Further, the normalization of the kinetic terms in the Lagrangian~\eqref{Meslagr}
provides
\begin{eqnarray}
V_1(0)=V_2(0)=\frac{F_{\pi}^2}{4}, \quad \quad V_4(0)= \frac{1}{12} (F_0^2-F_{\pi}^2).
\end{eqnarray}
In principle, the coupling of the $\eta_0$ singlet is different from $F_{\pi}$ because the
subgroup $U(3)_V$ does not present a nonet symmetry. However, in the large $N_c$-limit
$F_0=F_{\pi}$. Moreover, the quantity of $\bar{\theta}_0$ can be denoted in terms of
physical quantities \cite{Ottnad:2009jw}
\begin{equation}
  \bar{\theta}_0=\Bigg[1+ \frac{4 V_0^{(2)}}{F_{\pi}^2} \frac{4 M_K^2-M_{\pi}^2}{M_{\pi}^2
      (2M_{K}^2-M_{\pi}^2)}\Bigg]^{-1} \theta_0.
\end{equation}
Here, we note that $\bar{\theta}_0=\mathcal{O}(\delta ^2)$, and take $1/N_c=\mathcal{O}(\delta ^2)$
as counting rules \cite{Leutwyler:1996sa}. More detail and information on the formalism used
in the work can be found in e.g. in Refs.~\cite{Borasoy:2000pq, HerreraSiklody:1996pm}.

We now turn to the baryon sector of the effective Lagrangian.
In the SU(3) flavor representation the spin-$1/2$ anti-symmetric triplet and symmetric
sextet charmed baryon states are denoted as in the following matrices, respectively,
\[
  B_{\bar{3}}=
  \begin{pmatrix}
               0                    & \Lambda_{c}^+        & \Xi_{c}^+ \\
    - \Lambda_{c}^+      &        0                       & \Xi_{c}^0 \\
    - \Xi_{c}^+                & -\Xi_{c}^0                &     0
  \end{pmatrix},\quad
  B_6=
  \begin{pmatrix}
    \Sigma_{c}^{++}                           & \frac{\Sigma_{c}^+}{\sqrt{2}}           & \frac{\Xi_{c}^{'+}}{\sqrt{2}}  \\
    \frac{\Sigma_{c}^{+}}{\sqrt{2}}    & \Sigma_{c}^0                                  & \frac{\Xi_{c}^{'0}}{\sqrt{2}}  \\
    \frac{\Xi_{c}^{'+}}{\sqrt{2}}           & \frac{\Xi_{c}^{'0}} {\sqrt{2}}           & \Omega_{c}^{0}
  \end{pmatrix}.
\]
Similarly to the mesonic Lagrangian one can write down the most general effective Lagrangian
for the charmed baryon multiplets. Here, we only present the terms pertinent to the calculation.
%For a detail explanation, see e.g.~\cite{Borasoy:2000pq}.
In the quark mass and momentum expansion, the relevant free and interaction Lagrangians up to the second chiral order are given by \cite{Yan:1992gz,Borasoy:2000pq,Guo:2012vf,Jiang:2014ena,Xiabng:2018qsd},
\begin{equation}
\begin{aligned}
\mathcal{L}_{\phi B,\text{free}}^{(1)}= ~&\frac{1}{2} \langle \bar{B}_{\bar{3}}(i \slashed{D}-m_{\bar{3}})
B_{\bar{3}} \rangle+\langle \bar{B}_{6}(i \slashed{D}-m_{6}) B_{\bar{6}} \rangle,\\
\mathcal{L}_{ \phi B,\text{int}}=          ~ &g_1\langle \bar{B}_{6}\slashed{u} \gamma_5 {B}_{6}\rangle
+g_2[\langle \bar{B}_{6}\slashed{u} \gamma_5 {B}_{\bar{3}} \rangle+h.c.]+g_6\langle \bar{B}_{\bar{3}}
\slashed{u} \gamma_5 {B}_{\bar{3}} \rangle\ \\
&+g_1\langle \bar{B}_{6}\gamma^{\mu} \gamma_5 {B}_{6} \rangle \langle u_{\mu} \rangle
+g_2[\langle \bar{B}_{6}\gamma^{\mu} \gamma_5 {B}_{\bar{3}} \rangle+h.c.]  \langle u_{\mu} \rangle        +g_6\langle \bar{B}_{\bar{3}} \gamma^{\mu} \gamma_5  {B}_{\bar{3}} \rangle  \langle u_{\mu} \rangle,\\     \mathcal{L}_{\bar{3}\bar{3}}^{(2)}=    ~&w_{16/17} \langle \bar{B}_{\bar{3}} \sigma^{\mu \nu}F_{\mu \nu}^+
B_{\bar{3}} \rangle+w_{18} \langle \bar{B}_{\bar{3}} \sigma^{\mu \nu} B_{\bar{3}} \rangle 
\langle F_{\mu \nu}^+\rangle +b_{D/F} \langle \bar{B}_{\bar{3}} \tilde{\chi}_+ B_{\bar{3}} \rangle
+b_0 \langle \bar{B}_{\bar{3}} B_{\bar{3}} \rangle \langle \tilde{\chi}_+ \rangle \\
&+i w_{10/11} \frac{\sqrt{6}}{F_0}\eta_0 \langle \bar{B}_{\bar{3}} \tilde{\chi}_{-} B_{\bar{3}}
\rangle+i w_{12} \frac{\sqrt{6}}{F_0}\eta_0 \langle \bar{B}_{\bar{3}}  B_{\bar{3}} \rangle
\langle  \tilde{\chi}_{-} \rangle+i (w_{13/14}^{'} \bar{\theta}_0+w_{13/14}
\frac{\sqrt{6}}{F_0}\eta_0)  \langle  \bar{B}_{\bar{3}} \sigma^{\mu \nu} \gamma_5 F_{\mu \nu}^+
B_{\bar{3}} \rangle \\
&+i (w_{15}^{'} \bar{\theta}_0+w_{15}\frac{\sqrt{6}}{F_0}\eta_0) \langle
\bar{B}_{\bar{3}} \sigma^{\mu \nu} \gamma_5 B_{\bar{3}} \rangle \langle F_{\mu \nu}^+ \rangle, \\
\mathcal{L}_{66}^{(2)}= ~&w_{16/17} \langle \bar{B}_{6} \sigma^{\mu \nu} F_{\mu \nu}^+B_{6}
\rangle+w_{18}\langle \bar{B}_{6} \sigma^{\mu \nu} B_{6} \rangle \langle F_{\mu \nu}^+
\rangle +b_{D/F} \langle \bar{B}_{6} \tilde{\chi}_+ B_{6} \rangle + b_0 \langle \bar{B}_{6} B_{6}
\rangle \langle \tilde{\chi}_+ \rangle\\                  	                                     &+iw_{10/11} \frac{\sqrt{6}}{F_0}\eta_0 \langle \bar{B}_{6} \tilde{\chi}_{-} B_{6} \rangle
+i w_{12} \frac{\sqrt{6}}{F_0}\eta_0 \langle \bar{B}_{6}  B_{6} \rangle  \langle
\tilde{\chi}_{-} \rangle + i (w_{13/14}^{'} \bar{\theta}_0+w_{13/14}
\frac{\sqrt{6}}{F_0}\eta_0) \langle  \bar{B}_{6} \sigma^{\mu \nu} \gamma_5 F_{\mu \nu}^+ B_{6} \rangle \\
&+i  (w_{15}^{'} \bar{\theta}_0+w_{15} \frac{\sqrt{6}}{F_0}\eta_0) \langle
\bar{B}_{6} \sigma^{\mu \nu} \gamma_5 B_{6} \rangle  \langle F_{\mu \nu}^+ \rangle ,\\
\mathcal{L}_{6\bar{3}}^{(2)}=  ~& w_{16/17} \langle \bar{B}_{6} \sigma^{\mu \nu} F_{\mu \nu}^+
B_{\bar{3}} \rangle + w_{18} \langle B_{6} \sigma^{\mu \nu} B_{\bar{3}}\rangle \langle F_{\mu \nu}^+
\rangle+  b_{D/F} \langle \bar{B}_{6} \tilde{\chi}_+ B_{\bar{3}} \rangle +b_0 \langle \bar{B}_{6}
B_{\bar{3}} \rangle \langle \tilde{\chi}_+ \rangle~,     
\end{aligned}
\end{equation}
where the relevant building blocks are
%
%$\tilde{\chi}_{-}=~\chi_{-}-i\mathcal{A}(U+U^\dagger)-i \mathcal{B}\chi_{+}$,~~~ $\tilde{\chi}_{+}=~\chi_{+}-i\ma%thcal{A}(U-U^\dagger)-i \mathcal{B}\chi_{-}$, and the relevant building blocks are
%
\begin{equation}
  \begin{aligned}
  \tilde{\chi}_{-}= &\chi_{-}-i\mathcal{A}(U+U^\dagger)-i \mathcal{B}\chi_{+}~, \\
  \tilde{\chi}_{+}= &\chi_{+}-i\mathcal{A}(U-U^\dagger)-i \mathcal{B}\chi_{-}~, \\  
  D_{\mu} B= &\; \partial_\mu B + \Gamma_\mu B + B \Gamma_{\mu}^T~, \\
  \Gamma_{\mu}=&\;\frac{1}{2}[u^{\dag}(\partial{\mu}-ir_{\mu})u+u(\partial{\mu}-il_{\mu})u^{\dag}]~, \\ 
  u=&\;i[u^{\dag}(\partial{\mu}-ir_{\mu})u-u(\partial{\mu}-il_{\mu})u^{\dag}]~. \\
\end{aligned}
\end{equation}
The charge matrix for the singly-charmed baryons is $Q_h=\text{diag}(1, 0, 0)$, while for
the light quarks the charge matrix is $Q_l=\text{diag}(2/3, -1/3, -1/3)$. We use
$w_{10/11} + 3 w_{12}=w'_{10}$ as in Ref.~\cite{Ottnad:2009jw}.

As can be seen from the contributing Lagrangians, there are quite number of low-energy
constants (LECs). The meson-baryon coupling constants $g_i~(i=1,\ldots,6)$, the symmetry-breaking
LECs $b_D$ and $b_F$ as well as the LECs $w_{16/17}, w_{18}$ related to the
CP-conserving electromagnetic response can all be taken from earlier studies of different
observables, as detailed in Section~\ref{sec:res}.

This leaves us with the yet undetermined LECs $w'_{10}, w'_{13/14}, w'_{15}$
and $ w_{13/14}, w_{15}$. As will be shown, we
can fix $w_{13/14}, w_{15}$ from recent lattice results QCD for the neutron and
proton electric dipole moments, $d_n$ and $d_p$, respectively.
The remaining of these LECs will be varied as  $0^{+0.5}_{-0.5}$~GeV$^{-1}$, that is within a natural
range. This naive dimensional analysis should be eventually overcome by a more
sophisticated modeling of the LECs or invoking further lattice QCD results.
%All other LECs that appear can
%be fixed from magnetic moments, masses and other observables as detailed below.
Having fixed/estimated all the LECs will then allow to 
estimate the CP-violating contributions to the singly-charmed baryons induced by the
$\theta$-term.

\section{CP-violating electromagnetic form factor}
\label{sec:ff}
The electromagnetic form factors of a baryon are defined via the matrix element of
the electromagnetic current,
\begin{equation}
\begin{aligned}
  \Braket{B(p_{f})|J_ \text{em}^{\mu}|B(p_i)}
  = &\;\bar{u}(p_{f})\Big[\gamma^{\mu} F_1(q^2)-\frac{i F_2(q^2)}{2m_B} \sigma^{\mu \nu} q_{\nu} \\
     & + i (\gamma^{\mu} q^2 \gamma_5-2 m_B q^{\mu} \gamma_5) F_A(q^2)- \frac{F_3(q^2)}{2 m_B}
        \sigma^{\mu \nu} q_{\nu} \gamma_5 \Big]u(p_i)~,
\end{aligned}
\label{ff}
\end{equation}
with $q^2=(p_f-p_i)^2$  the invariant momentum transfer squared, $m_B$ the baryon
mass and $J_ \text{em}^{\mu}$ the electromagnetic current. Here, $F_1(q^2)$ and $F_2(q^2)$
are the P- and CP-conserving Dirac and Pauli form factors, respectively. $F_A(q^2)$
denotes the P-violating anapole form factor, and $F_3(q^2)$, which will be considered
throughout this work, the P- and CP-violating electric dipole form factor. The electric
dipole moment of the baryon $B$ is then given by
\begin{equation}
d_B = \frac{F_{3,B}(0)}{2m_B}~.
\end{equation}

\begin{figure}[htb!]
\centering
\includegraphics[width=0.8\textwidth]{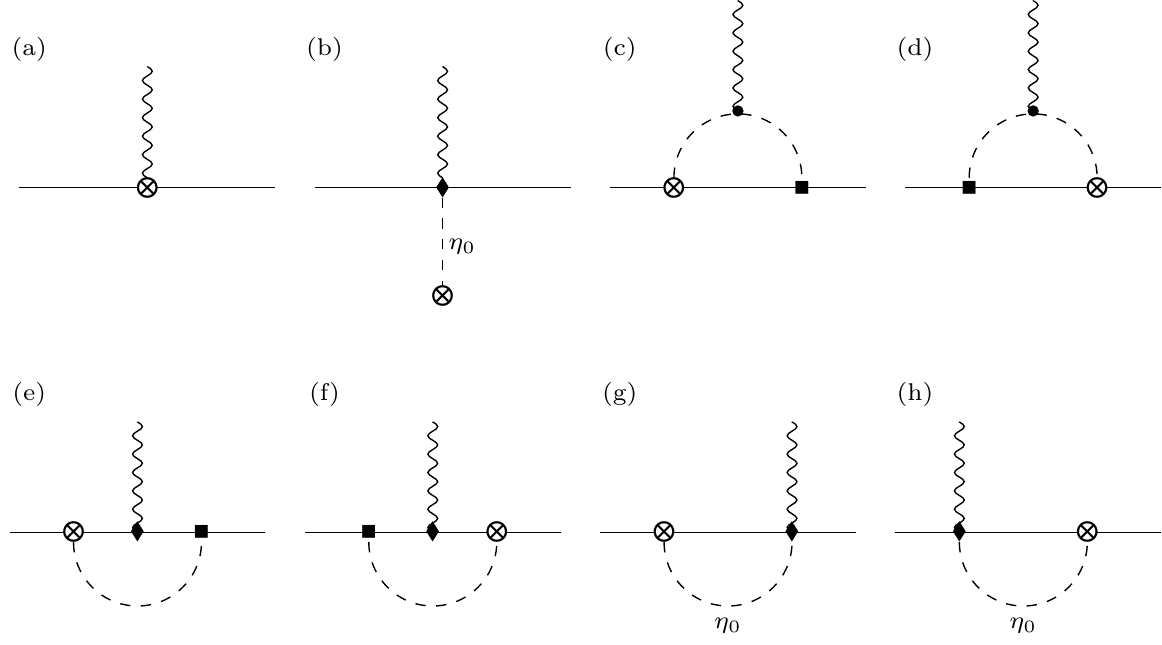}
\caption{CP-violating contributions of the spin-$1/2$ charmed baryons. Solid lines correspond
to contribution from either spin-$1/2$ anti-triplet or sextet multiplets of charmed baryons.
Filled circles are second-order mesonic vertices, squares and diamonds represent vertices
generated by the first and second order meson-baryon Lagrangian, respectively.
CP-violating vertices are denoted by $\otimes$.}
\label{fig:diag}
\end{figure}

In what follows, we will use the effective Lagrangian to calculate the CP-violating
form factor of the singly-charmed baryons at next-to-leading (NLO) order, which
includes tree as well as loop diagrams as shown in Figure~\ref{fig:diag}, where we
display the corresponding Feynman diagrams. Tree-level diagrams at leading order
are presented in (a) and (b). One-loop diagrams at order $\mathcal{O}(\delta^2)$ and
$\mathcal{O}(\delta^3)$  in (c)-(d), and (e)-(h), respectively.
The type of diagrams in (g)-(h) with pionic or kaonic loops of the antitriplet and
the sextet charmed baryons are cancelling each other, thus they are  not displayed here.

\begin{figure}[htb!]
\centering
\includegraphics[width=0.8\textwidth]{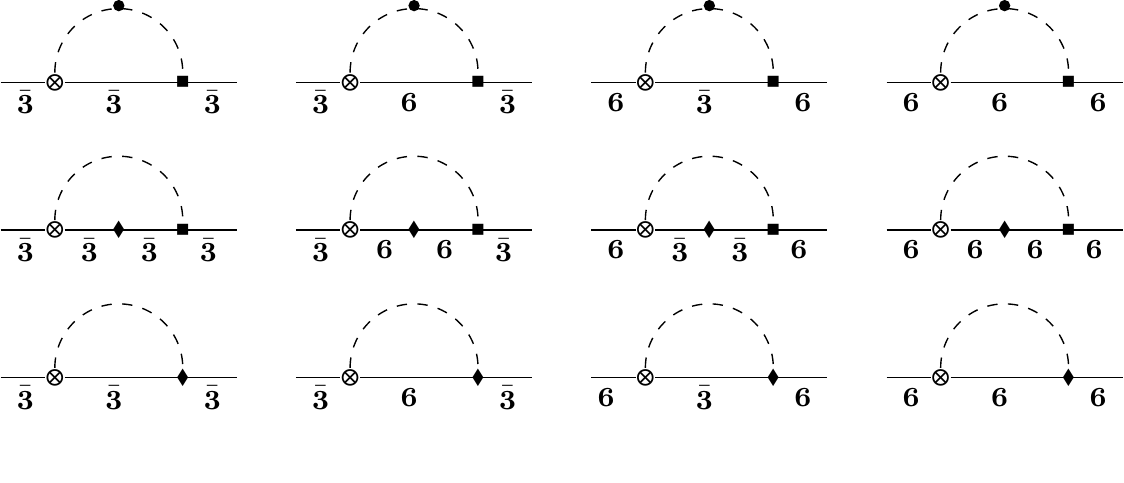}
\caption{Different combinations of the spin-$1/2$ anti-triplet and sextet charmed
baryons contributing to $F_3(q^2)$.}
\label{fig:Combi}
\end{figure}

We show different combinations of the charmed baryon states from anti-triplet and sextet multiplets considered throughout the calculation in Figure~\ref{fig:Combi}.

The results obtained for the form factor $F_3(q^2)$ of the charmed baryons coming
from the tree-level diagrams are collected in Table~\ref{tab:tree} with
\begin{equation}
\alpha = \frac{576 V_0^{(2)} V_3^{(1)}}{(F_0 F_{\pi} M_{\eta_{0}})^2}~.
\end{equation}
\begin{table}[ht!]
	\caption{Tree-level contribution to the $F_3(q^2)$ of the charmed baryons.}
	\centering
	\begin{tabular}{lll}
		\toprule
		%\multicolumn{2}{c}{States}                   \\
		%\cmidrule(r){1-2}
		States                &                                     & Contributions \\
		\midrule  
		                           & $\Lambda_{c}^+$      & $ e \bar{\theta}_{0}~m_{\Lambda_{c}}~[2 \alpha(w_{13/14} + 2 w_{15}) + 8 (w'_{13/14}+2   
		                                                                            w'_{15})]$ \\
		$B_{\bar{3}}$    & $\Xi_{c}^+$                & $ e \bar{\theta}_{0}~m_{\Xi_{c}}~[2 \alpha(w_{13/14} + 2 w_{15}) + 8 (w'_{13/14} + 2   
		                                                                            w'_{15})]$ \\
		                           & $\Xi_{c}^0$                & $ e \bar{\theta}_{0}~m_{\Xi_{c}}~4 (\alpha w_{15} + 4 w'_{15})$ \\
		\bottomrule
		                          & $\Sigma_{c}^{++}$    & $ e \bar{\theta}_{0}~m_{\Sigma_{c}}~[2\alpha(w_{13/14} + w_{15}) + 8 (w'_{13/14} +   
		                                                                              w'_{15})]$ \\
		                          & $\Sigma_{c}^{+}$       & $ e \bar{\theta}_{0}~m_{\Sigma_{c}}~[\alpha(w_{13/14} + 2 w_{15}) + 4 (w'_{13/14} + 2   
		                                                                              w'_{15})]$ \\
		                          & $\Sigma_{c}^{0}$       & $ e \bar{\theta}_{0}~m_{\Sigma_{c}}~2 (\alpha w_{15} + 4 w'_{15})$ \\
		$B_{6}$            & $\Xi_{c}^{'+}$              & $ e \bar{\theta}_{0}~m_{\Xi_{c}^{'}}~[\alpha(w_{13/14} + 2 w_{15}) + 4 (w'_{13/14}+2   
		                                                                           w'_{15})]$ \\
		                          & $\Xi_{c}^{'0}$              & $ e \bar{\theta}_{0}~m_{\Xi_{c}^{'}}~2 (\alpha w_{15} + 4 w'_{15})$ \\
		                          & $\Omega_{c}^{0}$      &$ e \bar{\theta}_{0}~m_{\Omega_{c}}~2 (\alpha w_{15} + 4 w'_{15})$ \\
		\bottomrule
	\end{tabular}
	\label{tab:tree}
\end{table}

As usual in the EOMS scheme, the loop contributions are rather lengthy expression. Let us
discuss the case of the $\Lambda_c^+$. The one-loop contribution can be 
be written as, cf. Fig.~\ref{fig:diag},
\begin{eqnarray}
\label{eq:loop}
F_{3,\Lambda_c^+}^{\rm loop} (q^2)
&=& \sum_{i=1}^2 \frac{ e \bar{\theta}_0 V_0^{(2)} \tilde{m}} { \pi^2 F_{\pi}^4} \frac{1} {(4 \tilde{m}^2-q^2)}
\Bigg[ C_{cd}^i \Big( m_i (\tilde{m}+ m_i) \Big) \Big( 2 J^{cd}_i (\tilde{m}^2,  m_i^2, M^2)- 2 J^{cd}_i (q^2, M^2, M^2) \nonumber\\
&+& (2 M^2+2\tilde{m}^2-2m_i^2-q^2) J^{cd}_i (\tilde{m}^2, \tilde{m}^2, q^2, M^2,  m_i^2, M^2) \Big)    \Bigg] \nonumber\\
&+&  \sum_{i=3}^{4} \frac{ e \bar{\theta}_0 V_0^{(2)}} { \pi^2 F_{\pi}^4} \frac{1} {(4 \tilde{m}^2-q^2)} 
          \Bigg[C_{ef}^i \Big(-J_i^{ef}(M_i^2) (4 \tilde{m}^2-q^2) + J_i^{ef}(\tilde{m}^2) (4 \tilde{m}^2-q^2) \nonumber
\end{eqnarray}
\begin{eqnarray}
&-& (M_i^2(q^2+4 \tilde{m}^2)-4 \tilde{m}^2 q^2) J^{ef}_i(\tilde{m}^2, \tilde{m}^2, M_i^2)+ 8 \tilde{m}^2 (M_i^2-2\tilde{m}^2) J^{ef}_i(q^2, \tilde{m}^2, \tilde{m}^2) \nonumber\\
&+& 4 \tilde{m}^2 M_i^2 (2 M_i^2-8 \tilde{m}^2+q^2) J^{ef}_i (\tilde{m}^2, \tilde{m}^2, q^2, \tilde{m}^2, M^2, \tilde{m}^2) \Big) \Bigg] \nonumber\\
&+&  \sum_{i=5}^{7} \frac{ e \bar{\theta}_0 V_0^{(2)}} { \pi^2 F_{\pi}^4} \frac{1} {(4 \tilde{m}^2-q^2)} 
          \Bigg[C_{ef}^i \Big(-J_i^{ef}(M_i^2) (4 \tilde{m}^2-q^2) + J_i^{ef}(m_i^2) (4 \tilde{m}^2-q^2) \nonumber\\
&-& \big(M_i^2(4 \tilde{m} m_i+q^2)+(\tilde{m+m_i})(4 \tilde{m}^2 m_i-m_i q^2-\tilde{m}(4 m_i^2+q^2))\big) J_i^{ef} (\tilde{m}^2, m_i^2 , M_i^2) \nonumber\\
&-& 4 \tilde{m} (\tilde{m}+m_i) (\tilde{m}^2+m_i^2-M_i^2) J_i^{ef} (q^2, m_i^2, m_i^2) + 2 \tilde{m} (\tilde{m}+m_i)  \nonumber\\
            &\times & \big(2 M_i^4 +(\tilde{m}^2-m_i^2)(2 \tilde{m}^2-2 m_i^2-q^2)+M_i^2 (q^2-4\tilde{m}^2-4 m_i^2)\big) J_i^{ef} (\tilde{m}^2, \tilde{m}^2, q^2, m_i^2, M_i^2, m_i^2 ) \Big) \Bigg] \nonumber
\end{eqnarray}
\begin{eqnarray}
&+&  \sum_{i=8}^{10} \frac{ e \bar{\theta}_0 V_0^{(2)}} { \pi^2 F_{\pi}^4} \frac{1} {(4 \tilde{m}^2-q^2)} 
     \Bigg[C_{ef}^i \Big(-J_i^{ef}(M_i^2) (4 \tilde{m}^2-q^2) + J_i^{ef}(m_i^2) (4 \tilde{m}^2-q^2) \nonumber\\
&+& \big(\tilde{m}^2 q^2-4\tilde{m}^3 m_i+q^2(m_i^2-M_i^2)+2 \tilde{m}m_i (q^2+2 m_i^2-2 M_i^2)\big)
            J_i^{ef}(\tilde{m}^2, m_i^2 , M_i^2) \nonumber\\
            &-& 4 \tilde{m} (\tilde{m}+m_i) (\tilde{m}^2+m_i^2-M_i^2) J_i^{ef} (q^2, m_i^2, m_i^2)
            + 2 \tilde{m} (\tilde{m}+m_i)  \nonumber\\
            &\times & \big(2 M_i^4 +2 \tilde{m}^4+M_i^2(q^2-4 m_i^2)+m_i^2(2 m_i^2+q^2)-\tilde{m}^2(4 M_i^2+4m_i^2+q^2))\big)
            J_i^{ef} (\tilde{m}^2, \tilde{m}^2, q^2, m_i^2, M_i^2, m_i^2 ) \Big) \Bigg] \nonumber\\
&+& \frac{16 e \bar{\theta}_0 V_0^{(2)}}{\pi^2 F_{\pi}^2 F_0^2 }  \Bigg[C_{gh}^{11} \Big(J_{11}^{gh}(M^2)
         -J_{11}^{gh}(\tilde{m}^2)-(M^2 - 4 \tilde{m}^2) J_{11}^{gh} (\tilde{m}^2, M^2) \Big) \Bigg]~,
\end{eqnarray}  
with $m_i$, $\tilde{m}_i$ and $M_i$ denoting the masses of
the corresponding internal and external baryons and meson running in the loop, for notational
simplicity. In the case at hand, $\tilde{m} = m_{\Lambda_c^+}$.  The $J(m_i,M_i,q^2)$ functions
can be reduced to the scalar loop functions given in
Appendix~\ref{sec:AppendixA}, and the labels $cd$, $ef$ and $gh$ refer to the types of
diagrams shown in Fig.~\ref{fig:diag}.
The corresponding coefficients $C_{cd}, C_{ef}$ and $C_{gh}$ for the $\Lambda_c^+$
together with the intermediate meson-baryon states
are shown in Table~\ref{tab:Lamb}, the corresponding tables for the other particles
can be found in Appendix~\ref{App:loop}.
A \textsc{Mathematica} notebook with these loop
functions can be obtained from the first author of this paper.

\begin{table}[ht!]
  \caption{Loop contribution to the $F_3(q^2)$ of the $\Lambda_{c}^+$ baryon
    with $\beta = (b_{D/F}+b_0+3 w{'}_{10})$.}
	\centering
	\begin{tabular}{cccc}
		\toprule
		%\multicolumn{2}{c}{States}                   \\
		%\cmidrule(r){1-2}
 \text{Diagram type}  & number & meson-baryon state &   Coefficient \\
 \midrule
 (c), (d) &  1 & $\Xi_{c}^0, K^{\pm}$ & $2 g_6 b_{D/F} $ \\
         &  2 & $\Xi_{c}^{'0}, K^{\pm}$  & $ g_2 b_{D/F} $ \\
 \midrule 			                   
         & 3  & $\Lambda_{c}^{+}, \eta_{8}$ & $\frac{8}{3} g_6 b_{D/F} (w_{16/17}+2w_{18}) $ \\
         & 4  & $\Lambda_{c}^{+}, \eta_{0}$ & $\frac{32}{3} \beta  g_6 (w_{16/17}+2w_{18}) $\\
         & 5  & $\Xi_{c}^{+}, K^0$          & $ 4 g_6 b_{D/F} (w_{16/17}+2w_{18})$ \\
 (e), (f)& 6  & $\Xi_{c}^{'0}, K^{\pm}$      & $ 2 g_2 b_{D/F} w_{18}$ \\ 
         & 7  & $\Xi_{c}^{'+}, K^{0}$        &$ g_2 b_{D/F} (w_{16/17}+2w_{18}) $ \\
         & 8  & $\Sigma_{c}^{0}, \pi^{\pm}$  & $ 4 g_2 b_{D/F} w_{18}$ \\
         & 9  & $\Sigma_{c}^{+}, \pi^{0}$    & $ 2 g_2 b_{D/F} (w_{16/17}+2w_{18})$ \\
         & 10 & $\Sigma_{c}^{++}, \pi^{\pm}$  & $ 4 g_2 b_{D/F} (w_{16/17}+w_{18})$ \\
	\midrule   		                                                  
 (g), (h)& 11 & $\Lambda_{c}^+, \eta_0$     & $\beta (w_{13/14} + 2 w_{15}) $ \\	                                                  
   \bottomrule
	\end{tabular}
	\label{tab:Lamb}
\end{table}

\section{Results}
\label{sec:res}

First, we must fix parameters. The pion decay constant is taken as $F_{\pi}=92.2$~MeV.
In what follows, due to the lack of data from the charmed meson sector,
  we make recourse to the ground state baryon octet as much as possible to fix as
  many LECs as possible. While this is an approximation, we expect that we are estimating
  at least the right order of magnitude of the EDMs of the charmed baryons. Consequently,
two symmetry-breaking LECs in the baryon sector can be obtained from baryon mass splittings.
We use $b_D=-0.606 ~ \text{GeV}^{-1}$  and $b_F=-0.209 ~ \text{GeV}^{-1}$
\cite{Borasoy:1996bx,Guo:2012vf}.
The tree-level contributions can be expressed in terms of two independent
linear combinations of unknown LECs as $\alpha (w_{13/14} + 4 w'_{13/14})$ and
$\alpha w_{15} + 4 w'_{15}$, cf.  Table~\ref{tab:tree}. The loop contributions are
also dependent on unknown LECs, viz., $w'_{10}$, $w_{13/14}$ and $w_{15}$. The conventional
magnetic moment couplings, $w_{18}$ is taken equal to  $w_{16/17}=0.40$, determined from
fits to calculations to
baryon magnetic moments in~\cite{Kubis:2000aa,Meissner:1997hn}. 

Further, $V_0^{(2)}=-5 \times 10^{-4} \; \text{GeV}^4$ and $V_3^{(1)}=3.5 \times 10^{-4} \;
\text{GeV}^2$ are the values obtained from an analysis of  $\eta- \eta'$ mixing in U(3)
chiral perturbation theory \cite{HerreraSiklody:1997kd}. The various baryon-meson couplings
are taken from Refs.~\cite{Jiang:2014ena, Jiang:2015xqa}, $g_1=0.98$, $g_2=-0.60$, $g_3=0.85$,
and $g_4=1.04$. Because of the forbidden $B_{\bar{3}} B_{\bar{3}} \phi$-vertex, we have $g_6=0$.
We use the physical masses of the pertinent mesons and baryons running in the corresponding
loops, cf. Tables~4-11.

As the unknown LECs cannot be parameterized such a common constant as in \cite{Guo:2012vf},
since the combinations coming from different particles are different, they have to be
considered individually. Using the lattice data from \cite{Dragos:2019oxn} at physical pion mass, we
use the neutron dipole moment to fix $\beta w_{15}$  from the  $\Xi_c^0$
by comparing the loop contributions. With that,
we can use the proton electric dipole moment to determine $ \beta w_{13/14}$
from the $\Lambda_c^+$. 
We get
\begin{eqnarray}
   \beta w_{13/14}=  - 0.00435 ~ \text{GeV}^{-1},
  \qquad \beta w_{15}= 0.00175~ \text{GeV}^{-1},
\end{eqnarray}
With these obtained values, we take the variation of $w'_{10}, w'_{13/14}$ and
$w'_{15}$, and calculate the CP-violating form factor $F_3(q^2)$ for the
singly-charmed baryons in the range $q^2 \simeq 0.05 \ldots 0.3\,$GeV$^2$ as given in Table 12-14.
We are well aware that there are other determinations of $d_p$ and $d_n$ in the literature, see
e.g. Refs.~\cite{Shintani:2015vsx,Guo:2015tla},
and that there is an on-going debate on the axial rotation in the finite volume (mixing
between the form factors $F_2$ and $F_3$, see e.g. Ref.~\cite{Abusaif:2019gry}). However,
since our study is largely exploratory, we do not explore the whole
possible parameter space.

The electric dipole moments for the various baryons are collected in Table~\ref{tab:edm}.
As there is a sizeable uncertainty induced by the unknown LECs, we refrain from performing
a systematic error analysis accounting e.g. for the effects of higher orders in the
chiral expansion. 
Hopefully, lattice QCD will be able to supply pertinent information on the
LECs so that more accurate predictions can be made.
\renewcommand{\arraystretch}{1.3}
\begin{table}[htb!]
\caption{Electric dipole moments for the singly-charmed baryons in units of $e\,\theta_0\,\text{fm}$.  
Set~1,2,3 refers to $w'_{10} = w'_{13/14}= w'_{15} = -0.5, 0, +0.5$, in order.}
\begin{center}  
\begin{tabular}{cccccccccc}
\hline    
    & $\Lambda_c^{+}$ & $\Xi_c^{+} $ & $\Xi_c^{0} $ & $\Sigma_c^{++}$ & $\Sigma_c^{+}$
& $\Sigma_c^{0}$ & $\Xi_c^{'+}$ & $\Xi_c^{'0}$ & $\Omega_c^{0}$ \\ \hline
Set~1 & $0.0476$ &  $0.0470$ & $0.0294$ & $0.0047$ & $0.0053$ & $0.0058$ & $0.0007$ & $0.0085$ & $0.0067$\\
Set~2 & $0.0011$ &  $0.0005$ & $-0.0015$ & $-0.0090$ & $-0.0049$ & $-0.0010$ & $-0.0097$ & $0.0015$ & $-0.0003$\\
Set~3 & $-0.0454$ &  $-0.0460$ & $-0.0324$ & $-0.0228$ & $-0.0153$ & $-0.0078$ & $-0.0202$ & $-0.0053$ & $-0.0074$\\
\hline
\end{tabular}
\end{center}
\label{tab:edm}
\end{table}

\section{Conclusion}
\label{sec:summ}

In this paper, we have performed a one-loop calculation of the CP-violating form factor
$F_3(q^2)$ and the corresponding electric dipole moments of the spin-1/2 singly-charmed
baryons, where the mechanism of the CP violation is the QCD $\theta$-term. Not all the
appearing low-energy constants could be fixed from experimental or lattice QCD data,
so the resulting predictions show a spread, cf. Table~\ref{tab:edm}  and the tables
in Appendix~\ref{sec:AppendixB}. We hope that with more lattice QCD studies on strong
CP violations, these LECs can be determined and more accurate predictions can be made,
not to mention possible experimental determinations.

%%%%%%%%%%%%%%%%%%%%%%%%%%%%%%%%%%%%%%%%%%%%%%%%%%%%%%%%%%%%%%%%%%%%%%%%%%%%%%%%
%
%							ACKNOWLEDGMENTS
%
%%%%%%%%%%%%%%%%%%%%%%%%%%%%%%%%%%%%%%%%%%%%%%%%%%%%%%%%%%%%%%%%%%%%%%%%%%%%%%%%
\section*{Acknowledgments}
We thank Jambul Gegelia, Sebastian Neubert, Altu\u{g} Özpineci and Daniel Severt for discussions.
Partial financial support from the Deutsche Forschungsgemeinschaft (SFB/TRR~110, ``Symmetries
and the Emergence of Structure in QCD''), by the Chinese 
Academy of Sciences (CAS) President's International Fellowship Initiative (PIFI) (grant no. 2018DM0034) 
by VolkswagenStiftung (grant no. 93562) and by the EU (Strong2020) is acknowledged.

\newpage

%%%%%%%%%%%%%%%%%%%%%%%%%%%%%%%%%%%%%%%%%%%%%%%%%%%%%%%%%%%%%%%%%%%%%%%%%%%%%%%%
%
%							APPENDIX
%
%%%%%%%%%%%%%%%%%%%%%%%%%%%%%%%%%%%%%%%%%%%%%%%%%%%%%%%%%%%%%%%%%%%%%%%%%%%%%%%%
\begin{appendix}
\section{Scalar loop integrals}\label{sec:AppendixA}
The scalar loop integrals of one-, two-, and three-point functions which are used for the
calculation of the diagrams are given by
\begin{align*}
&J_0(m^2)= \frac{(2\pi\mu)^{4-d}}{i\pi^2} \int \frac{d^d k}{k^2-m^2+i0^+}, \\
&J_0(p^2, m_1^2, m_2^2)= \frac{(2\pi\mu)^{4-d}}{i\pi^2} \int \frac{d^d k}{[k^2-m_1^2+i0^+][(k+p)^2-m_2^2+i0^+]}, \\
&J_0(p_i^2, (p_f-p_i)^2, p_f^2, m_1^2, m_2^2, m_3^2) \\
&= \frac{(2\pi\mu)^{4-d}}{i\pi^2} \int \frac{d^d k}{[k^2-m_1^2+i0^+][(k-p_i)^2-m_2^2+i0^+][(k-p_f)^2-m_3^2+i0^+]}.
\end{align*}

\section{Loop contributions}
\label{App:loop}

All one-loop contributions to the various baryons take the form as given in Eq.~\ref{eq:loop}.
In this appendix, we collect the corresponding intermediate meson-baryon states and the
values of the coefficients $C^{cd}, C^{ef}$ and $C^{gh}$ for the baryons not given in the
main text.

\begin{table}[ht!]
	\caption{Loop contribution to the $F_3(q^2)$ of the $\Xi_{c}^+$ baryon.}
	\centering
	\begin{tabular}{cccc}
		\toprule
		%\multicolumn{2}{c}{States}                   \\
		%\cmidrule(r){1-2}
 \text{Diagram type}  & Number & Meson-baryon state &   Coefficient \\
    \midrule	                           
	                                & 1 &   $\Xi_{c}^0, \pi^{\pm}$                         & $ g_6 b_{D/F} $ \\
	 (c), (d)                   & 2 & $\Omega_{c}^{0}, K^{\pm}$                & $ g_2 b_{D/F} $ \\
                            		& 3 & $\Sigma_{c}^{++}, K^{\pm}$               & $ g_2 b_{D/F} $ \\
		                           & 4  & $\Xi_{c}^{'0}, \pi^{\pm}$                      & $ g_2 b_{D/F}  $\\
 \bottomrule	                                                  
                                    & 5  & $\Xi_{c}^{+}, \eta_8$                            & $ g_6 b_{D/F} (w_{16/17}+2w_{18})$ \\
	                                & 6  & $\Xi_{c}^{+}, \eta_{0}$                          & $  \beta g_6 (w_{16/17}+2w_{18})$ \\ 
                                    & 7  & $\Xi_{c}^{+}, \pi^{0}$                            & $ g_6 b_{D/F}  (w_{16/17}+2w_{18})$ \\
                                    & 8 & $\Xi_{c}^{'0}, \pi^{\pm}$                       &$ g_2 b_{D/F} w_{18} $ \\
	(e), (f)  	                  & 9 & $\Xi_{c}^{'+}, \eta_{8}$                         & $ g_2 b_{D/F} (w_{16/17}+2w_{18})$ \\
		                           & 10 & $\Sigma_{c}^{+}, K^{0}$                     & $ g_2 b_{D/F} (w_{16/17}+2w_{18})$ \\
		                           & 11 & $\Xi_{c}^{'+}, \pi^{0}$                         & $ g_2 b_{D/F} (w_{16/17}+2w_{18})$ \\
		                           & 12 & $\Sigma_{c}^{++}, K^{\pm}$              & $ g_2 b_{D/F} (w_{16/17}+w_{18})$ \\
		                           & 13 & $\Omega_{c}^{0}, K^{\pm}$                & $ g_2 b_{D/F} w_{18}$ \\                 
	\midrule      
  	  (g), (h)                  & 14 & $\Xi_{c}^+, \eta_0$                              & $\beta (w_{13/14} + 2 w_{15}) $ \\		                                                      
 \bottomrule
	\end{tabular}
	\label{tab:Zp}
\end{table}

\begin{table}[ht!]
	\caption{Loop contribution to the $F_3(q^2)$ of the $\Xi_{c}^0$ baryon.}
	\centering
	\begin{tabular}{cccc}
		\toprule
		%\multicolumn{2}{c}{States}                   \\
		%\cmidrule(r){1-2}
\text{Diagram type}  & Number & Meson-baryon state &   Coefficient \\
	\midrule	                           
	                                                  &1  & $\Lambda_{c}^+, K^{\pm}$                   & $ g_6 b_{D/F} $ \\
	 (c), (d)                                       &2  & $\Omega_{c}^{0}, K^{\pm}$                 & $ g_2 b_{D/F} $ \\
                            		                    & 3 & $\Xi_{c}^{+}, \pi^{\pm}$                       & $ g_6 b_{D/F} $ \\
		                                              & 4 & $\Sigma_{c}^{+}, K^{\pm}$                 & $ g_2 b_{D/F}  $\\
		                                                & 5 & $\Xi_{c}^{'+}, \pi^{\pm}$                     & $ g_2 b_{D/F}  $\\
 \bottomrule	                                                  
                                                        & 6  & $\Xi_{c}^{0}, \eta_8$                           & $ g_6 b_{D/F} w_{18}$ \\
	                                                     & 7 & $\Xi_{c}^{0}, \eta_{0}$                         & $ \beta g_6 w_{18} $ \\ 
	                                                      & 8 & $\Xi_{c}^{0}, \pi^{0}$                            & $ g_6 b_{D/F} w_{18}$ \\
	 (e), (f)      	                              & 9 & $\Xi_{c}^{'+}, \pi^{\pm}$                       &$ g_2 b_{D/F} (w_{16/17}+2w_{18}) $ \\
		                                                 & 10 & $\Sigma_{c}^{0}, K^{0}$                      & $ g_2 b_{D/F} w_{18}$ \\
		                                                  & 11 & $\Xi_{c}^{'0}, \eta_{8}$                      & $ g_2 b_{D/F} w_{18}$ \\
		                                                 & 12  & $\Xi_{c}^{'0}, \pi^{0}$                        & $ g_2 b_{D/F} w_{18}$ \\
		                                                 & 13  & $\Sigma_{c}^{+}, K^{\pm}$               & $ g_2 b_{D/F} (w_{16/17}+2w_{18})$ \\
		                                                 & 14 & $\Omega_{c}^{0}, K^{0}$                  & $ g_2 b_{D/F} w_{18}$ \\           
	\midrule      
  	  (g), (h)                                    & 15  & $\Xi_{c}^0, \eta_0$                                & $\beta w_{15} $ \\		                                                            
 \bottomrule
	\end{tabular}
	\label{tab:Z0}
\end{table}

\newpage

\begin{table}[ht!]
	\caption{Loop contribution to the $F_3(q^2)$ of the $\Sigma_{c}^{++}$ baryon.}
	\centering
	\begin{tabular}{cccc}
		\toprule
		%\multicolumn{2}{c}{States}                   \\
		%\cmidrule(r){1-2}
\text{Diagram type}  & Number & Meson-baryon state &   Coefficient \\
	\midrule	                           
	                                                   & 1 & $\Xi_{c}^{'+}, K^{\pm}$                      & $ g_1 b_{D/F} $ \\
	 (c), (d)                                         & 2 & $\Sigma_{c}^{+}, \pi^{\pm}$              & $ g_1 b_{D/F} $ \\
                            		                    & 3 & $\Xi_{c}^{+}, K^{\pm}$                       & $ g_2 b_{D/F} $ \\
		                                                & 4 & $\Lambda_{c}^{+}, \pi^{\pm}$           & $ g_2 b_{D/F}  $\\
 \bottomrule	                                                  
                                                       & 5 & $\Sigma_{c}^{++}, \eta_8$                 & $ g_1 b_{D/F} (w_{16/17}+w_{18}) $ \\
	                                                   & 6 & $\Sigma_{c}^{++}, \eta_{0}$               & $ \beta g_1 (w_{16/17}+w_{18}) $ \\ 
	 (e), (f)                                       & 7 & $\Sigma_{c}^{++}, \pi^{0}$                 & $ g_1 b_{D/F} (w_{16/17}+w_{18}) $ \\
		                                                & 8 & $\Xi_{c}^{+}, K^{\pm}$                       &$ g_2 b_{D/F} (w_{16/17}+2w_{18}) $ \\
		                                                & 9 & $\Lambda_{c}^{0}, \pi^{\pm}$           & $ g_2 b_{D/F} (w_{16/17}+2w_{18}) $ \\            
\midrule      
  	  (g), (h)                                      &10  & $\Sigma_{c}^{++}, \eta_0$                & $ \beta (w_{13/14} + w_{15}) $ \\		                                                  
 \bottomrule
	\end{tabular}
	\label{tab:Sigpp}
\end{table}

\begin{table}[ht!]
	\caption{Loop contribution to the $F_3(q^2)$ of the $\Sigma_{c}^{+}$ baryon.}
	\centering
	\begin{tabular}{cccc}
		\toprule
		%\multicolumn{2}{c}{States}                   \\
		%\cmidrule(r){1-2}
\text{Diagram type}  & Number & Meson-baryon state &   Coefficient \\
	\midrule	                           
      (c), (d)  	                                & 1 & $\Xi_{c}^{'0}, K^{\pm}$                      & $ g_1 b_{D/F} $ \\	                                                       
                            		                    & 2 & $\Xi_{c}^{0}, K^{\pm}$                       & $ g_2 b_{D/F} $ \\		                                                
   \bottomrule	                                                  
                                                        & 3 & $\Sigma_{c}^{+}, \eta_8$                  & $ g_1 b_{D/F} (w_{16/17}+2w_{18}) $ \\
	                                                    & 4 & $\Sigma_{c}^{+}, \eta_{0}$                & $ \beta g_1 (w_{16/17}+2w_{18}) $ \\ 
	 (e), (f)                                        & 5 & $\Xi_{c}^{+}, K^{0}$                           & $ g_2 b_{D/F} (w_{16/17}+w_{18}) $ \\
		                                                & 6 & $\Xi_{c}^{0}, K^{\pm}$                       &$ g_2 b_{D/F} w_{18} $ \\
		                                                & 7 & $\Lambda_{c}^{0}, \pi^{0}$               & $ g_2 b_{D/F} (w_{16/17}+2w_{18}) $ \\           
	\midrule      
  	  (g), (h)                                      & 8 & $\Sigma_{c}^{+}, \eta_0$                  & $  \beta (w_{13/14} +2 w_{15})$ \\		                                                  
   \bottomrule
	\end{tabular}
	\label{tab:Sigp}
\end{table}

\begin{table}[ht!]
	\caption{Loop contribution to the $F_3(q^2)$ of the $\Sigma_{c}^{0}$ baryon.}
	\centering
	\begin{tabular}{cccc}
		\toprule
		%\multicolumn{2}{c}{States}                   \\
		%\cmidrule(r){1-2}
\text{Diagram type}  & Number & Meson-baryon state &   Coefficient \\
	\midrule	                           
      (c), (d)  	                               & 1 & $\Sigma_{c}^{+}, \pi^{\pm}$             & $ g_1 b_{D/F} $ \\	                                                       
                            		                   & 2 & $\Lambda_{c}^{0}, \pi^{\pm}$          & $ g_2 b_{D/F} $ \\		                                                
 \bottomrule	                                                  
                                                         & 3 & $\Sigma_{c}^{0}, \eta_8$                 & $ g_1 b_{D/F} w_{18} $ \\
	                                                    & 4 & $\Sigma_{c}^{0}, \eta_{0}$               & $ \beta g_1 w_{18} $ \\ 
	 (e), (f)                                        & 5 & $\Sigma_{c}^{0}, \pi^{0}$                  & $ g_1 b_{D/F} w_{18} $ \\
		                                                & 6 & $\Xi_{c}^{0}, K^{0}$                          &$ g_2 b_{D/F} w_{18} $ \\
		                                                & 7  & $\Lambda_{c}^{0}, \pi^{\pm}$          & $ g_2 b_{D/F} (w_{16/17}+2w_{18}) $ \\     
\midrule      
  	  (g), (h)                                    & 8 & $\Sigma_{c}^{0}, \eta_0$                  & $\beta w_{15}  $ \\		                                                         
 \bottomrule
	\end{tabular}
	\label{tab:Sig0}
\end{table}

\newpage

\begin{table}[ht!]
	\caption{Loop contribution to the $F_3(q^2)$ of the $\Xi_{c}^{'+}$ baryon.}
	\centering
	\begin{tabular}{cccc}
		\toprule
		%\multicolumn{2}{c}{States}                   \\
		%\cmidrule(r){1-2}
\text{Diagram type}  & Number & Meson-baryon state &   Coefficient \\
  \midrule	                           
	                                                   & 1 & $\Omega_{c}^0, K^{\pm}$                      & $ g_1 b_{D/F} $ \\
	 (c), (d)                                        & 2 & $\Sigma_{c}^{++}, K^{\pm}$                  & $ g_1 b_{D/F} $ \\
                            		                      & 3 & $\Xi_{c}^{'0}, \pi^{\pm}$                         & $ g_1 b_{D/F} $ \\
		                                                & 4 & $\Xi_{c}^{0}, \pi^{\pm}$                        & $ g_2 b_{D/F}  $\\
 \bottomrule	                                                  
                                                        & 5 & $\Xi_{c}^{'+}, \eta_8$                            & $ g_1 b_{D/F} (w_{16/17}+2w_{18})$ \\
	                                                    & 6 & $\Xi_{c}^{'+}, \eta_{0}$                          & $ \beta g_1 (w_{16/17}+2w_{18}) $ \\ 
	                                                   & 7 & $\Xi_{c}^{'+}, \pi^{0}$                            & $ g_1 b_{D/F}  (w_{16/17}+2w_{18})$ \\
    (e), (f) 		                                  & 8 & $\Sigma_{c}^{+}, K^{0}$                       &$ g_1 b_{D/F} (w_{16/17}+2w_{18}) $ \\
		                                               & 9 & $\Xi_{c}^{+}, \eta_{8}$                          & $ g_2 b_{D/F} (w_{16/17}+2w_{18})$ \\
		                                                & 10 & $\Lambda_{c}^{+}, K^{0}$                   & $ g_2 b_{D/F} (w_{16/17}+2w_{18})$ \\
		                                                & 11 & $\Xi_{c}^{+}, \pi^{0}$                          & $ g_2 b_{D/F} (w_{16/17}+2w_{18})$ \\
		                                                 & 12 & $\Xi_{c}^{0}, \pi^{\pm}$                      & $ g_2 b_{D/F} w_{18}$ \\		    
\midrule      
  	  (g), (h)                                       & 13 & $\Xi_{c}^{'+}, \eta_0$                              & $\beta (w_{13/14} + 2 w_{15})$ \\		                 
\bottomrule
	\end{tabular}
	\label{tab:Zvp}
\end{table}

\begin{table}[ht!]
	\caption{Loop contribution to the $F_3(q^2)$ of the $\Xi_{c}^{'0}$ baryon.}
	\centering
	\begin{tabular}{cccc}
		\toprule
		%\multicolumn{2}{c}{States}                   \\
		%\cmidrule(r){1-2}
\text{Diagram type}  & Number & Meson-baryon state &   Coefficient \\
	\midrule	                           
	                                                    & 1  & $\Sigma_{c}^{+}, K^{\pm}$                     & $ g_1 b_{D/F} $ \\
	 (c), (d)                                        & 2 & $\Xi_{c}^{'+}, \pi^{\pm}$                        & $ g_1 b_{D/F} $ \\
                            		                     & 3 & $\Lambda_{c}^{+}, K^{\pm}$                & $ g_2 b_{D/F} $ \\
		                                                & 4 & $\Xi_{c}^{+}, \pi^{\pm}$                         & $ g_2 b_{D/F}  $\\		                                                  		\bottomrule	                                                  
                                                         & 5 & $\Xi_{c}^{'0}, \eta_8$                           & $ g_1 b_{D/F} w_{18}$ \\
	                                                    & 6  & $\Xi_{c}^{'0}, \eta_{0}$                         & $ \beta g_1 w_{18} $ \\ 
	                                                    & 7 & $\Xi_{c}^{'0}, \pi^{0}$                            & $ g_1 b_{D/F} w_{18}$ \\
	  (e), (f)  		                             & 8  & $\Sigma_{c}^{0}, K^{0}$                       &$ g_1 b_{D/F} w_{18}$ \\
  	                                                     & 9 & $\Omega_{c}^{0}, K^{0}$                      & $ g_1 b_{D/F} w_{18}$ \\
		                                               & 10 & $\Lambda_{c}^{0}, K^{\pm}$                & $ g_2 b_{D/F} (w_{16/17}+2w_{18})$ \\
		                                              & 11  & $\Xi_{c}^{0}, \eta_{8}$                          & $ g_2 b_{D/F} w_{18}$ \\
		                                               & 12   & $\Xi_{c}^{0}, \pi^{0}$                             & $ g_2 b_{D/F} w_{18}$ \\		  
     \midrule       
  	  (g), (h)                                       & 13 & $\Xi_{c}^{'0}, \eta_0$                                & $\beta w_{15} $ \\		                                                     \bottomrule
	\end{tabular}
	\label{tab:Zv0}
\end{table}

\begin{table}[ht!]
	\caption{Loop contribution to the $F_3(q^2)$ of the $\Omega_{c}^{0}$ baryon.}
	\centering
	\begin{tabular}{cccc}
		\toprule
		%\multicolumn{2}{c}{States}                   \\
		%\cmidrule(r){1-2}
\text{Diagram type}  & Number & Meson-baryon state &   Coefficient \\
     \midrule	                           
      (c), (d)  	               & 1 & $\Xi_{c}^{'+}, K^{\pm}$                        & $ g_1 b_{D/F} $ \\	                                                       
                            		 & 2 & $\Xi_{c}^{+}, K^{\pm}$                        & $ g_2 b_{D/F} $ \\		                                                
     \bottomrule	                                                  
                                    & 3 & $\Omega_{c}^{0}, \eta_8$                  & $ g_1 b_{D/F} w_{18} $ \\
    (e), (f)   	             & 4  & $\Omega_{c}^{0}, \eta_{0}$                & $ \beta g_1 w_{18} $ \\ 
	                               & 5 & $\Xi_{c}^{+}, K^{\pm}$                         & $ g_2 b_{D/F} (w_{16/17}+2w_{18}) $ \\
		                           & 6& $\Xi_{c}^{0}, K^{0}$                              &$ g_2 b_{D/F} w_{18} $ \\		
    \midrule      
  	(g), (h)                   & 7 & $\Omega_{c}^{0}, \eta_0$                  & $  \beta w_{15} $ \\		                                                                                                              	\bottomrule
	\end{tabular}
	\label{tab:Om}
\end{table}

\newpage

\section{Results for the CP-violating form factor}\label{sec:AppendixB}

This appendix contains results for the loop contributions to the form factor $F_3(q^2)$ for the various baryons, for photon
virtualities below $q^2 \simeq 0.3\,$GeV$^2$.

\begin{table}[ht!]
	\caption{Loop contribution to the $F_3(q^2)$ of the $B_{\bar{3}}$ and $B_6$ states for $w'_{10}, w'_{13/14}, w'_{15} = -0.5$.}
	\centering
	\begin{tabular}{@{}lccccccccc@{}}
		\toprule
$q^2 (\text{GeV}^2)$ & $\Lambda_c^{+}$ & $\Xi_c^{+} $ & $\Xi_c^{0} $ & $\Sigma_c^{++}$ & $\Sigma_c^{+}$ & $\Sigma_c^{0}$ & $\Xi_c^{'+}$ & $\Xi_c^{'0}$ & $\Omega_c^{0}$ \\
     \bottomrule			                             
    \bottomrule			                                                                                                                                
$0.0484$ & $1.1028$ & $1.1772$ & $0.7362$ & $-0.1150$ & $0.1339$ & $0.3832$& $0.0361$ & $0.2102$ & $0.1855$ \\
$0.1024$ & $1.1015$ & $1.1764$ & $0.7339$ & $-0.0434$ & $0.1359$ & $0.3164$ & $0.0452$ & $0.2047$ & $0.1861$ \\
$0.1444$ & $1.1004$ & $1.1756$ & $0.7323$ & $-0.0175$ & $0.1374$ & $0.2943$ & $0.0500$ & $0.2027$ & $0.1865$ \\
$0.1936$ & $1.0992$ & $1.1747$ & $0.7305$ & $0.0021$ & $0.1392$ & $0.2789$ & $0.0546$ & $0.2014$ & $0.1871$ \\
$0.2500$ & $1.0978$ & $1.1735$ & $0.7285$ & $0.0181$ & $0.1412$ & $0.2679$ & $0.0590$ & $0.2007$ & $0.1877$ \\
$0.3136$ & $1.0963$ & $1.1722$ &  $0.7264$ & $0.0319$ & $0.1435$ & $0.2598$ & $0.0633$ & $0.2006$ & $0.1885$ \\                                       
    \bottomrule	
	\end{tabular}
	\label{tab:m1}
\end{table}

\begin{table}[ht!]
	\caption{Loop contribution to the $F_3(q^2)$ of the $B_{\bar{3}}$ and $B_6$ states for $w'_{10}, w'_{13/14}, w'_{15} = 0$}
	\centering
	\begin{tabular}{@{}lccccccccc@{}}
    \toprule
$q^2 (\text{GeV}^2)$ & $\Lambda_c^{+}$ & $\Xi_c^{+}$ & $\Xi_c^{0} $ & $\Sigma_c^{++}$ & $\Sigma_c^{+}$  & $\Sigma_c^{0}$ & $\Xi_c^{'+} $ & $\Xi_c^{'0}$ & $\Omega_c^{0}$\\
   \bottomrule		        
   \bottomrule	                                                                                                                                                       
$0.0484$ & $0.0245$ & $0.0121$ & $-0.0414$ & $-0.4594$ & $-0.1246$ & $0.2102$& $-0.2403$ & $0.0253$ & $-0.0108$ \\
$0.1024$ & $0.0232$ & $0.0113$ & $-0.0437$ & $-0.3912$ & $-0.1252$ & $0.1417$ & $-0.2339$ & $0.0181$ & $-0.0121$ \\
$0.1444$ & $0.0221$ & $0.0105$ & $-0.0453$ & $-0.3680$ & $-0.1257$ & $0.1183$ & $-0.2311$ & $0.0147$ & $-0.0130$ \\
$0.1936$ & $0.0209$ & $0.0096$ & $-0.0471$ & $-0.3513$ & $-0.1262$ & $0.1015$ & $-0.2289$ & $0.0118$ & $-0.0140$ \\
$0.2500$ & $0.0195$ & $0.0085$ & $-0.0491$ & $-0.3388$ & $-0.1268$ & $0.0887$ & $-0.2272$ & $0.0094$ & $-0.0152$ \\
$0.3136$ & $0.0180$ & $0.0071$ &  $-0.0512$ & $-0.3290$ & $-0.1275$ & $0.0786$ & $-0.2259$ & $0.0073$ & $-0.0165$ \\                                       
 	\bottomrule		 	
	\end{tabular}
	\label{tab:zer}
\end{table}

\begin{table}[ht!]
	\caption{Loop contribution to the $F_3(q^2)$ of the $B_{\bar{3}}$ and $B_6$ states for $w'_{10}, w'_{13/14}, w'_{15} = 0.5$.}
	\centering
	\begin{tabular}{@{}lccccccccc@{}}              
     \toprule                        
$q^2 (\text{GeV}^2)$ & $\Lambda_c^{+}$ & $\Xi_c^{+}$ & $\Xi_c^{0} $ & $\Sigma_c^{++}$ & $\Sigma_c^{+}$ & $\Sigma_c^{0}$ & $\Xi_c^{'+} $ & $\Xi_c^{'0}$ & $\Omega_c^{0}$ \\
   \bottomrule	
   \bottomrule	 	                                                                                                                                                       
$0.0484$ & $-1.0545$ & $-1.1538$ & $-0.8155$ & $-0.8065$ & $-0.3837$ & $0.0391$& $-0.5173$ & $-0.1575$ & $-0.2053$ \\
$0.1024$ & $-1.0558$ & $-1.1546$ & $-0.8177$ & $-0.7416$ & $-0.3869$ & $-0.0310$ & $-0.5135$ & $-0.1665$ & $-0.2083$ \\
$0.1444$ & $-1.0569$ & $-1.1553$ & $-0.8194$ & $-0.7211$ & $-0.3893$ & $-0.0557$ & $-0.5127$ & $-0.1712$ & $-0.2106$ \\
$0.1936$ & $-1.0581$ & $-1.1563$ & $-0.8212$ & $-0.7075$ & $-0.3921$ & $-0.0741$ & $-0.5129$ & $-0.1757$ & $-0.2133$ \\
$0.2500$ & $-1.0595$ & $-1.1574$ & $-0.8231$ & $-0.6986$ & $-0.3954$ & $-0.0887$ & $-0.5139$ & $-0.1799$ & $-0.2163$ \\
$0.3136$ & $-1.0610$ & $-1.1588$ &  $-0.8253$ & $-0.6927$ & $-0.3991$ & $-0.1008$ & $-0.5156$ & $-0.1841$ & $-0.2196$ \\                                       
    \bottomrule	   	   	
	\end{tabular}
	\label{tab:p1}
\end{table}

\end{appendix}

\newpage

%%–––––––––––––––––––––––––––––––––––
%% Literature sources 
%%–––––––––––––––––––––––––––––––––––

{}
%%=============================================  
\end{document}